\documentclass[conference]{IEEEtran}
\usepackage{amssymb}
\usepackage{graphics}
\usepackage{epsfig}
\usepackage{mathptmx}
\usepackage{times}
\usepackage{amsmath}
\usepackage{amssymb}
\usepackage{graphicx}
\usepackage{amssymb}
\usepackage{epstopdf}
\usepackage{subfig}
\usepackage{float}
\usepackage{multicol}
\usepackage{multirow}
\usepackage{amsthm}
\usepackage{setspace}
\usepackage{amssymb}
\usepackage{amsfonts}%

\newtheorem{theorem}{Theorem}
\newtheorem{remark}{Remark}

\newtheorem{definition}{Definition}
\newtheorem{example}{Example}

\ifCLASSINFOpdf
\else
\fi

\begin{document}
%
\title{Fractional Order Hybrid Systems and Their Stability}

\author{S. Hassan HosseinNia, In\'es Tejado and Blas M. Vinagre
\IEEEauthorblockA{\\Department of Electrical, Electronic and Automation Engineering\\ 
Industrial Engineering School, University of Extremadura, 06006 Badajoz, Spain\\
  e-mail: \{hoseinnia;itejbal;bvinagre\}@unex.es}}

\maketitle

\begin{abstract}
This paper deals with hybrid systems (HS) with fractional order dynamics and their stability. The stability of two particular types of fractional order hybrid systems (FOHS), i.e., switching and reset control systems, is studied. Common Lyapunov method, as well as its frequency domain equivalence, are generalized for the former systems and, for the latter, H$_{\beta}$-condition is used --frequency domain equivalence of Lyapunov-like method for reset control systems. The applicability and efficiency of the proposed methods are shown by some illustrative examples.

\end{abstract}


\IEEEpeerreviewmaketitle

\section{Introduction}

Hybrid systems (HS) are heterogeneous dynamic systems whose behaviour is determined by interacting continuous-variable and discrete-event dynamics, and they arise from the use of finite-state logic to govern continuous physical processes or from topological and networks constraints interacting with continuous control \cite{Gollu_89,Shorten1996,schumacher_99,Goebel_09}. Typically, their stability is analyzed by Lyapunov's theory (see e.g.~\cite{Liberzon03,Narendra_94,Mori_98,Shim_98}. However, recently a frequency domain method equivalent to the common Lyapunov was proposed in \cite{Kunze08} to analyze the stability of a particular class of HS.

This paper deals with HS with fractional order dynamics and summarizes our stability results proposed in \cite{Hosseinnia_12b,Hosseinnia13a,Hosseinnia13b} for two types of HS: fractional order switching and reset control systems. In \cite{Hosseinnia_12c,Hosseinnia2013} the applied the proposed stability analysis to design a robust fractional-order controller. The motivation for studying switching systems came partly from the fact that such systems have numerous applications in control of mechanical systems, process control, automotive industry, power systems, traffic control, and so on, as well as there exists a large group of nonlinear systems which can be stabilized by switching control schemes, but not by any continuous static state feedback control law \cite{Lin09,Hespanha_04}. Likewise, reset control systems arise from overcoming the limitations in linear systems. On the other hand, fractional dynamics can be found in control systems due to both the system itself and the used control strategy (see e.g. \cite{Monje10,Podlubny_99a}). As a result, studying fractional dynamics of HS may be also an interesting topic.

The rest of the paper is organized as follows. Section \ref{sec_pre} gives some basic theorems and definitions concerning stability of both fractional and integer order systems. In Section \ref{sec_fos}, stability conditions for fractional order switching systems are established. Section \ref{sec_reset} addresses stability of fractional order reset control systems. Section \ref{sec_examples} gives some examples to show the applicability and goodness of the developed stability theory. Finally, Section \ref{sec_conclu} draws the concluding remarks.

\section{Preliminaries}
\label{sec_pre}

This section recalls some basic definitions and theorems concerning stability of fractional and integer order systems which will be useful to present the results of the following sections.

\subsection{Definitions}

Switching systems are hybrid dynamical systems consisting of a family of continuous-time subsystems
and a rule that orchestrates the switching among them \cite{Liberzon03,Daafouz02}. On the other hand, reset control systems are a class of HS \cite{banos2011} which include a linear controller which resets some of their states to zero when their input is zero or certain non-zero values.
A fractional order linear time invariant (FO-LTI) system can be given by:
\begin{equation}
D^{\alpha}x=Ax, x\in\mathbb{R}^{n}\label{FOLTI}%
\end{equation}
where $\alpha$ is the fractional order and the operator $D^{\alpha}$ denotes Riemann-Liouville definition given by (see e.g.~\cite{Podlubny_99a}):
\begin{eqnarray}
_aD^ \alpha f(t)= \frac{1}{\Gamma(1-\alpha)}\frac{d}{dx}\int_a^t{\frac{f(t)}{\left(t-\tau\right)}d\tau}
\label{RL}
\end{eqnarray}
where $e$ and $t$ are the lower and upper bounds of the operation, $\alpha \in\mathbb{R}$ is the order and $[.]$ means the integer part.

\subsection{Stability of fractional order systems}

\begin{theorem}
[\cite{Moze_07}] A fractional order system (\ref{FOLTI}) with order $\alpha$,
$0<\alpha\leq1$, is $t^{-a }$ asymptotically stable if and
only if there exists a positive definite matrix $P\in\mathbb{R}^{n}$ such that
\begin{equation}
\mathcal{A}  ^{T}P+P \mathcal{A}  <0,
\end{equation}
where $\mathcal{A}= -\left(  -A\right)  ^{\frac{1}{2-\alpha}}$.
\label{Moze0}
\end{theorem}

\begin{theorem}
[\cite{Moze_07}] A fractional system (\ref{FOLTI}) with order $\alpha$, $1
\leq\alpha< 2$, is $t^{-a}$ asymptotically stable if and only if there exists a
matrix $P=P^{T} > 0$, $P \in\mathbb{R}^{n \times n}$, such that
\begin{equation}
\label{FSQ}%
\begin{bmatrix}
\left(  A^{T}P+PA \right) \sin\phi  & \left(
A^{T}P-PA \right) \cos\phi \\
\left(  -A^{T}P+PA \right) \cos\phi  & \left(
A^{T}P+PA \right) \sin\phi
\end{bmatrix}<0,
\end{equation}
where $\phi=\frac{\alpha\pi}{2}$.
\label{Moze}
\end{theorem}

\begin{theorem}
[\cite{Boyd_94}] A system given by (\ref{SWHM}) is quadratically stable if
and only if there exists a matrix $P=P^{T} > 0$, $P \in\mathbb{R}^{n\times
n}$, such that
\[
\label{SWST}A_{i}^{T}P+PA_{i}<0, \forall i=1, ..., L.
\]
\end{theorem}


\begin{theorem} [\cite{Kunze08}] Consider $c_{1}(s)$ and $c_{2}(s)$, two stable polynomials of order $n$ corresponding to the subsystems $\dot{x}=A_{1}x$ and $\dot{x}=A_{2}x$, respectively, then the following statements are equivalent:

\begin{enumerate}
\item {$\frac{c_{1}(s)}{c_{2}(s)}$ and $\frac{c_{2}(s)}{c_{1}(s)}$ are strictly positive real (SPR).}

\item {$\left|  \arg(c_{1}(j\omega)) - \arg(c_{2}(j\omega))\right|  < \frac{\pi}{2}$, $\forall$ $\omega$.}

\item { $A_{1}$ and $A_{2}$ are quadratically stable, which means that $\exists P
=P^{T} >0 \in\mathbb{R}^{n\times n}$ such that $A_{1}^{T}P+PA_{1} <0$ ,
$A_{2}^{T}P+PA_{2} <0$.}
\end{enumerate}
\label{Freq_stab}
\end{theorem}

\subsection{Stability of integer order switching systems}

Consider a switching system as follows:
\begin{equation}
\dot{x}=Ax, A \in co\left\{ A_{1}, ..., A_{L} \right\} ,
\label{Conv}
\end{equation}
where $co$ denotes the convex combination and $A_{i}$, $i=1,...,L$, is the switching subsystem, which can be alternatively written as \cite{Pardalos_87}:
\begin{equation}
\label{SWHM}\dot{x}=Ax, A=\sum_{i=1}^{L} \lambda_{i} A_{i}, \forall\lambda_{i}
\geq0, \sum_{i=1}^{L} \lambda_{i}=1.
\end{equation}

\begin{theorem}[Lyapunov-like theorem \protect\cite{Goebel_09}\label{HLS}]
Consider a closed-loop reset system given by (\ref{CLeq}). If there exists a
Lyapunov-function candidate $V(x)$ such that
\begin{eqnarray}
\dot{V}(x)<0, \ x(t) \notin  \mathcal{M},
\label{LLC1}
\end{eqnarray}
\begin{eqnarray}
\bigtriangleup V(x)=V(x(t^+))-V(x(t))\leq0, \ x(t) \in  \mathcal{M},
\label{LLC2}
\end{eqnarray}
then there exists a left-continuous function $x(t)$ satisfying (\ref{CLeq}) for all $t\geq0$,
and the equilibrium point $x_e$ is globally uniformly asymptotically stable.
\label{TLL}
\end{theorem}

\subsection{Dynamics of reset control systems \label{Dreset}}

The dynamics of a reset controller can be described by a FDI equation as:
\begin{eqnarray}
\begin{matrix}
D^\alpha x_r(t)=A_r x_r(t)+B_re(t), \ e(t)\neq0,\\ 
x_r(t^+)=A_{R_r}x_r(t), \ e(t)=0,\\
u_r(t)=C_rx_r(t),
\end{matrix}
\label{reseteq}
\end{eqnarray}
where $0< \alpha \leq 1$ is the order of differentiation, $x_r(t) \in \mathbb{R}^{nr}$ is the reset controller state and $u_r(t) \in \mathbb{R}$ is its output. The matrix $A_{R_r}  \in \mathbb{R}^{n_r\times n_r}$ identifies that subset of states $x_r$ that are reset (the last ${\mathcal{R}}$ states) and has the form $A_{R_r}=\begin{bmatrix}
I_{n_{\bar{\mathcal{R}}}}& 0 \\ 
0 & 0_{n_{\mathcal{R}}}
\end{bmatrix}$ with $n_{\bar{\mathcal{R}}}=n_r-n_{\mathcal{R}}$. The linear controller $C(s)$ and plant $P(s)$ have, respectively, state space representations as follows:
\begin{eqnarray}
\begin{matrix}
D^\alpha x_c(t)=A_cx_c(t)+B_cu_r(t),\\ 
u_c(t)=C_cx_c(t),
\end{matrix}
\label{Conteq}
\end{eqnarray}
and
\begin{eqnarray}
\begin{matrix}
D^\alpha x_p(t)=A_px_p(t)+B_pu_c(t),\\ 
y(t)=C_px_p(t),
\end{matrix}
\label{Syseq}
\end{eqnarray}
where $A_p\in \mathbb{R}^{n_p \times n_p}$, $B_p \in \mathbb{R}^{n_p\times 1}$, $C_p \in \mathbb{R}^{1\times n_p}$, $A_c\in \mathbb{R}^{n_c\times n_c}$, $B_c \in \mathbb{R}^{n_c\times 1}$ and $C_c \in \mathbb{R}^{1\times n_c}$. Therefore, the closed-loop reset control system can be then described by the following FDI:
\begin{eqnarray}
\begin{matrix}
D^\alpha x(t)=A_{cl} x(t)+B_{cl}r, \ x(t) \notin  \mathcal{M}\\ 
x(t^+)=A_{R}x(t), \  x(t)\in \mathcal{M}\\
y(t)=C_{cl}x(t)
\end{matrix}
\label{CLeq}
\end{eqnarray}
where $t^+$ denotes $t+k$ and $k$ is sample time, 
$x=\begin{bmatrix}
x_p\\ 
x_c\\ 
x_r
\end{bmatrix}$,
$A_{cl}=\begin{bmatrix}
 A_p & B_pC_c & 0 \\ 
 0& A_c & B_cC_r \\ 
 -B_rC_p& 0 & A_r
\end{bmatrix}$,
$A_{R}=\begin{bmatrix}
 I_{n_p} & 0 & 0 \\ 
 0 & I_{n_c}& 0 \\ 
 0 & 0 & A_{R_r}
\end{bmatrix}$, $C_{cl}=\begin{bmatrix}
C_p & 0 & 0 \end{bmatrix}$ and $B_{cl}=\begin{bmatrix}
0 & 0 & B_r \end{bmatrix}^T$. The reset surface $\mathcal{M}$ is defined by:
\begin{eqnarray}
\begin{matrix}
\mathcal{M}=\left \{ x\in\mathbb{R}^n: C_{cl}x=0,\ (I-A_R)x\neq0 \right \},
\end{matrix}
\label{surf}
\end{eqnarray}
where $n=n_r+n_c+n_p$.

\section{Stability of Fractional Order Switching Systems}
\label{sec_fos}

Our objective hereafter is to establish stability conditions for fractional order switching systems. In this section, we firstly present the stability of such systems by common Lyapunov functions, which have been previously generalized to fractional order switching systems, and further its equivalence in frequency domain. This developed theory can be found in \cite{Hosseinnia_12b,Hosseinnia13a}.

\subsection{Common Lyapunov theory}

Consider a fractional order switching system of the form of (\ref{Conv}) as 
\begin{equation}
D^{\alpha}{x}=Ax, A \in co\left\{ A_{1}, ..., A_{L} \right\}.
\label{FSWHM}
\end{equation}

\begin{theorem}
A fractional system described by (\ref{FSWHM}) with order $\alpha$, $1 \leq\alpha< 2$, is stable if and only if there exists a matrix $P=P^{T} > 0$, $P
\in\mathbb{R}^{n \times n}$, such that
\small
\begin{align}%
\begin{bmatrix}
\left(  A_{i}^{T}P+PA_{i} \right) \sin \phi  &
\left(  A_{i}^{T}P-PA_{i} \right) \cos \phi  \\
\left(  -A_{i}^{T}P+PA_{i} \right) \cos \phi &
\left(  A_{i}^{T}P+PA_{i} \right) \sin \phi
\end{bmatrix}
<0,
\forall i=1,..., L,
\end{align}
\normalsize
where $\phi=\frac{\alpha\pi}{2}$.
\label{FSQ2}
\end{theorem}

\begin{theorem}
A fractional system given by (\ref{FSWHM}) with order $\alpha$, $0<\alpha\leq1$, is stable if and only if there exists a matrix $P=P^{T} > 0$, $P
\in\mathbb{R}^{n \times n}$, such that
\begin{equation}
  \mathcal{A}_i^{T}P+P\mathcal{A}_i  <0,\ \ \forall i=1, ..., L.
\end{equation}
\label{FSQ00}
\end{theorem}

\subsection{Frequency domain approach}

Next, frequency domain stability conditions will be given for fractional order switching systems based on results in \cite{Kunze08}. Consider a stable pseudo-polynomial of order $n\alpha$ of system (\ref{FSWHM}) as
\begin{equation}
d(s)=s^{n\alpha}+d_{n-1}s^{(n-1)\alpha}+ \cdots+d_{1}s^{\alpha}+d_{0},
\end{equation}
and a polynomial of order $n$ of system $\dot{\tilde{x}}=\tilde{A}\tilde{x}$ as
\begin{equation}
c(s)=s^{n}+c_{n-1}s^{(n-1)}+ \cdots+c_{1}s +c_{0}. \label{TCP}%
\end{equation}

In the following, the necessary and sufficient condition for the
stability for fractional order switching systems will be given.

\begin{theorem}
Consider $d_{1}(s)$ and $d_{2}(s)$, two stable pseudo-polynomials of order
$n$ corresponding to the subsystems $D^{\alpha}x=A_{1}x$ and
$D^{\alpha}x=A_{2}x$ with order $\alpha$, $1\leq\alpha< 2$, respectively, then the following
statements are equivalent:

\begin{enumerate}
\item {%
$ \left|  \arg\left( \det((A_{1}^{2}-\omega^{2}I)-2j \omega A_{1}\sin \phi) \right) - \right. \\ \left. \arg\left(  \det((A_{2}^{2}-\omega^{2}I)-2j \omega A_{2}\sin \phi)\right) \right|  < \frac{\pi}{2}, \forall\omega
$,\\ being $I$ the identity matrix with proper dimensions.}

\item { $A_{1}$ and $A_{2}$ are stable, which means that $\exists P
=P^{T} >0 \in\mathbb{R}^{n\times n}$ such that
\begin{align*}%
\begin{bmatrix}
\left(  A_{i}^{T}P+PA_{i} \right) \sin \phi & \left(
A_{i}^{T}P-PA_{i} \right) \cos \phi\\
\left(  -A_{i}^{T}P+PA_{i} \right) \cos \phi & \left(
A_{i}^{T}P+PA_{i} \right) \sin \phi
\end{bmatrix}
<0,
\forall i=1, 2.
\end{align*}
}
\end{enumerate}

\label{Freq_stab_frac}
\end{theorem}

\begin{theorem}
Consider two stable fractional order subsystems $D^{\alpha}x=A_{1}x$ and
$D^{\alpha}x=A_{2}x$ with order $\alpha$, $0<\alpha\leq1$, then the following
statements are equivalent:

\begin{enumerate}
\item {$\left|  \arg(\det(\mathcal{A}_{1} -j\omega I) )-\arg(\det
(\mathcal{A}_{2} -j\omega I))\right|  < \frac{\pi}{2}$, $\forall$ $\omega$.}

\item {$A_{1}$ and $A_{2}$ are stable, which means that $\exists P
=P^{T} >0 \in\mathbb{R}^{n\times n}$ such that }
\[
  \mathcal{A}_{i}   ^{T}P+P\mathcal{A}_{i}  <0, \forall i=1, 2.
\]
\end{enumerate}
 \label{Freq_stab_frac0}
\end{theorem}

Although the theory developed in the frequency domain does not necessarily prove the SPRness, a relation equivalent to the stability was obtained.

\section{Stability of Fractional Order Reset Control Systems}
\label{sec_reset}

In this section, fractional reset control systems will be handled by fractional order differential inclusion (FDI) equations.  Stability of this kind of systems will be analyzed using Lyapunov-like method presented previously. This developed theory can be found in \cite{Hosseinnia13b}.

\begin{definition} Reset control system (\ref{CLeq}) is said to satisfy the H$_\beta$-condition if there exists a $\beta \in \mathbb{R}^{n_{\mathcal{R}}}$ and a positive-definite matrix $P_{\mathcal{R}} \in \mathbb{R}^{n_{\mathcal{R}} \times n_{\mathcal{R}}}$ such that
\begin{eqnarray}
H_\beta(s)=\begin{bmatrix}
\beta C_p & 0_{n_{\bar{\mathcal{R}}}} & P_{\mathcal{R}}
\end{bmatrix}\left ( sI-\mathcal{A} \right )^{-1}\begin{bmatrix}
0\\ 
0^T_{\bar{\mathcal{R}}}\\ 
I_{\mathcal{R}}
\end{bmatrix},
\label{Hb}
\end{eqnarray}
where $\mathcal{A}= \left( -\left(-A_{cl}\right)  ^{\frac{1}{2-\alpha}}\right)$.
\end{definition} 

According to \cite{beker1999,beker2004,chen2000}, an integer-order reset control system of the form of (\ref{CLeq}) --with $\alpha=1$-- is asymptotically stable if and only if it satisfies the H$_\beta$-condition. The same idea can be used to prove the stability of fractional-order reset systems.

Now, consider $V(z(t))=z(t)^T\mathcal{P}z(t),\ \mathcal{P}\in\mathbb{R}^{N\times N}$ as a Lyapunov candidate for the unforced reset system (\ref{CLeq}) ($r=0$) where $x=[0,\cdots,0,1]z(t),\ z(t)\in\mathbb{R}^{N\times N}, \ \dot{z}=A_{f}z(t),$ and \small $A_{f}=
\begin{bmatrix}
0 & \cdots & 0 & A^{1/\alpha}\\
A^{1/\alpha} & \cdots & 0 & 0\\
& \ddots &  & \vdots\\
0 &  & A^{1/\alpha} & 0
\end{bmatrix}$ \normalsize (see \cite{Moze_07} more details for this transformation). Then, in accordance with \cite{Moze_07}, the necessary and sufficient condition to satisfy $\dot{V}(z(t))<0$ when $\frac{2}{3}<\alpha\leq1$ is:
$\left(A^\frac{1}{\alpha}\right) ^{T}P+P\left(A^\frac{1}{\alpha}\right) <0, \ x(t) \notin \mathcal{M},$
where $P(\subset\mathcal{P})\in\mathbb{R}^{n\times n}>0$. Likewise, based on results stated in Theorem \ref{Moze0}, the necessary and sufficient condition for $0<\alpha\leq1$ is
\begin{eqnarray}
\nonumber
\mathcal{A} ^{T}P+P\mathcal{A}  <0, \ x(t) \notin \mathcal{M}.
\label{unf}
\end{eqnarray}
Transforming the second equation of reset system (\ref{CLeq}), we have
\begin{eqnarray}
z(t^+)=\begin{bmatrix}
I_{N-n}& 0\\
0 & A_R
\end{bmatrix}z(t),
\label{disnew}
\end{eqnarray}
where $I_{N-n}$ is identity matrix with dimension of ${N-n}$. Thus, $\bigtriangleup V(z(t))<0$ if 
\begin{equation}
z^T(t)\left(\begin{bmatrix}
I_{N-n}& 0\\
0 & A^T_R
\end{bmatrix}\mathcal{P}+\mathcal{P}\begin{bmatrix}
I_{N-n}& 0\\
0 & A_R
\end{bmatrix}\right)z(t)<0.
\label{deltaz}
\end{equation}
Then, (\ref{deltaz}) is satisfied if $V(x(t^+))-V(x(t))$,
\begin{equation*}
x^T(t)(A_R^TPA_R-P\leq0)x(t)\leq0, \ x(t) \in  \mathcal{M}.
\end{equation*}
Therefore, Theorem \ref{TLL} can be reshaped in the following remark.

\begin{remark}
\label{remarkres1}
Choosing $V(z)=z(t)^T\mathcal{P}z(t),\ \mathcal{P}\in\mathbb{R}^{N\times N}$ as a Lyapunov candidate, and applying Theorem \ref{Moze0}, fractional-order reset system (\ref{CLeq}) is asymptotically stable if and only if:
\begin{eqnarray}
\label{Fcr1}
\mathcal{A} ^{T}P+P\mathcal{A}  <0, \ x(t) \notin  \mathcal{M},\\
\label{Fcr2}
A_R^TPA_R-P\leq0, \ x(t) \in  \mathcal{M}.
\end{eqnarray} 
\end{remark}
Remark \ref{remarkres1} can be also applicable transforming x to $\bold{x}(t)=x(t)-x_e=x(t)+A_{cl}^{-1}B_{cl}r$ in (\ref{CLeq}) in a special case of the reset control system with constant input $r$.
Define $\tilde{\mathcal{M}}=\left \{ x\in\mathbb{R}^n: C_{cl}x(t)=r \right \}$, and let $\Phi$ be a matrix whose columns span $\tilde{\mathcal{M}}$. Since $\tilde{\mathcal{M}} \subset  \mathcal{M}$, (\ref{Fcr2}) is implied by
\begin{eqnarray}
\label{Fcr}
\Phi \left (A_R^TPA_R-P<0 \right ) \Phi\leq0.
\end{eqnarray}

A straightforward computation shows that inequality (\ref{Fcr}) holds for some positive-definite symmetric matrix $P$ if there exists a $\beta \in \mathbb{R}^{n_{\mathcal{R}}}$ and a positive-definite $P_{\mathcal{R}} \in \mathbb{R}^{n_{\mathcal{R}} \times n_{\mathcal{R}}} $ such that
\begin{eqnarray}
\begin{bmatrix}
0 & 0_{\bar{\mathcal{R}}} & I_{\mathcal{R}}
\end{bmatrix} P
=\begin{bmatrix} \beta C_p & 0_{n_{\bar{\mathcal{R}}}} & P_{\mathcal{R}}
\end{bmatrix}.
\label{Hbeta}
\end{eqnarray}
To analyze stability, it suffices to find a positive-definite symmetric matrix $P$ such that (\ref{Fcr1}) and (\ref{Hbeta}) hold. Taking into account Kalman-Yakubovich-Popov (KYP) lemma \cite{slotine1991}, such $P$ exists if H$_\beta(s)$ in (\ref{Hb}) is strictly positive real (SPR) for some $\beta$. In addition, in accordance with \cite{ioannou1987}, it is obvious that the H$_\beta(s)$ is SPR if 
\begin{eqnarray}
\label{FrSPR}
\left |\arg(H_\beta(j\omega)) \right |<\frac{\pi}{2}, \forall \omega.
\end{eqnarray}
Therefore, these results can be stated in the following theorem.

\begin{theorem} The closed-loop fractional-order reset control system (\ref{CLeq}) is asymptotically stable if and only if it satisfies the H$_\beta$-condition (\ref{Hb}) or its phase equivalence (\ref{FrSPR}).
\label{FRQS}
\end{theorem}

\section{Examples}
\label{sec_examples}

This section gives some examples in order to show the applicability and effectiveness of the stability theories developed for FOHS. To this respect, phase portraits and time responses of the systems will be shown.

\begin{figure}[ptbh]
\begin{center}
\includegraphics[width=0.45\textwidth]{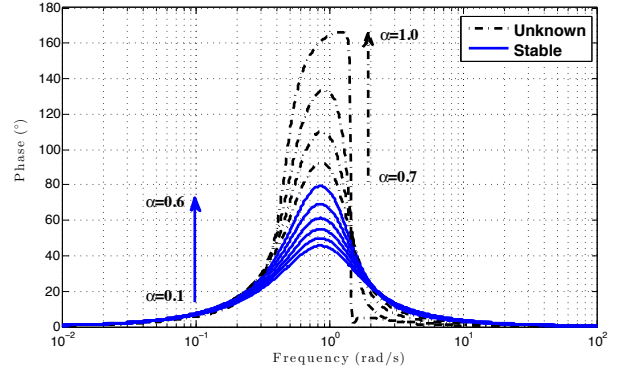}
\end{center}
\caption{Phase differences of characteristic polynomials of system in Example~\ref{expg} for different values of its order $\alpha$, $0<\alpha \leq 1$}%
\label{Expgsw}%
\end{figure}

\begin{figure}[ptbh]
\begin{center}
\includegraphics[width=0.35\textwidth]{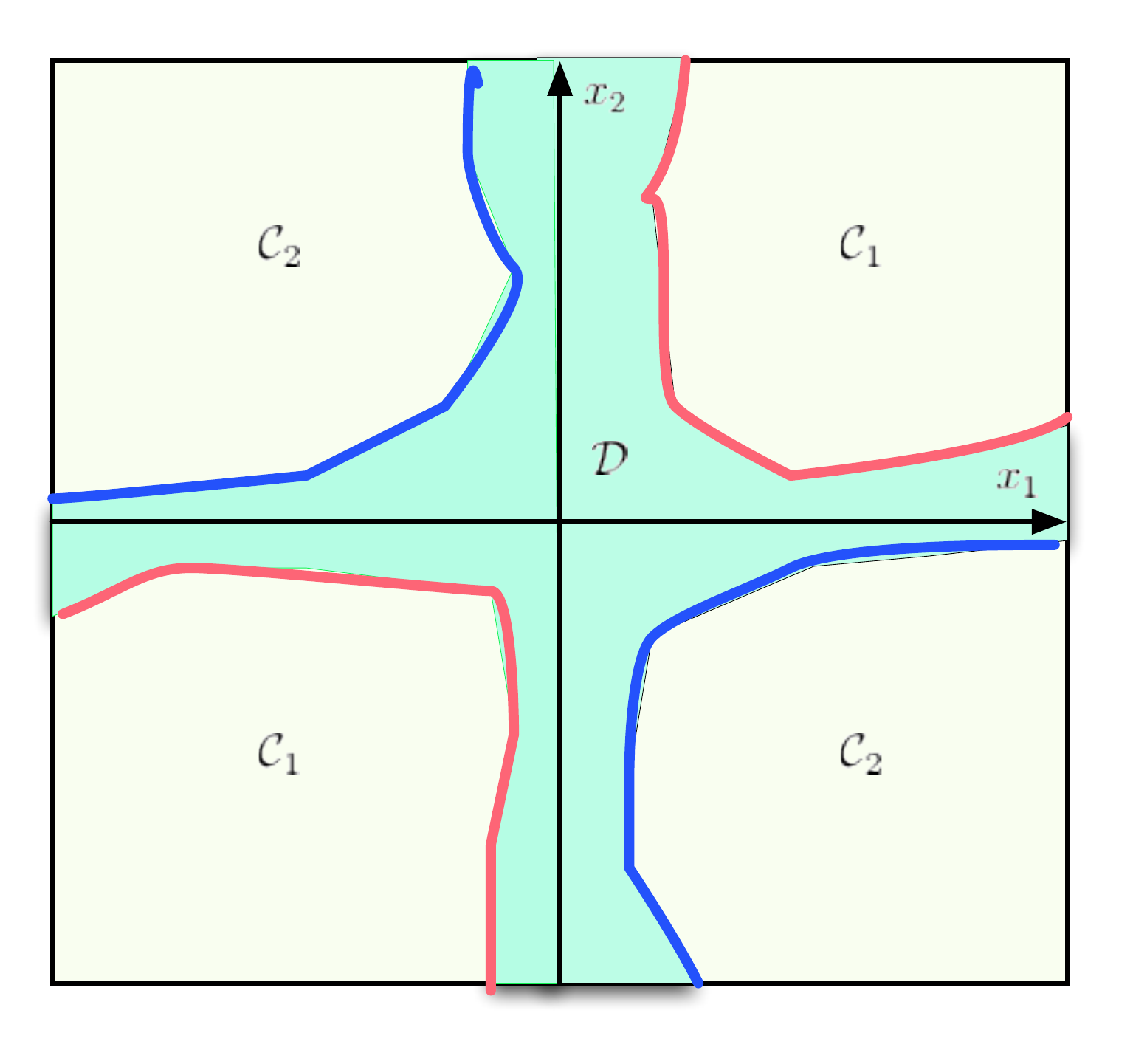}
\end{center}
\caption{Switching region for random switching of system in Example~\ref{expg}}%
\label{SZ}%
\end{figure}

\begin{figure}[ht]
\centering
\begin{tabular}{c}
($a$)\\
   \includegraphics[width=0.4\textwidth] {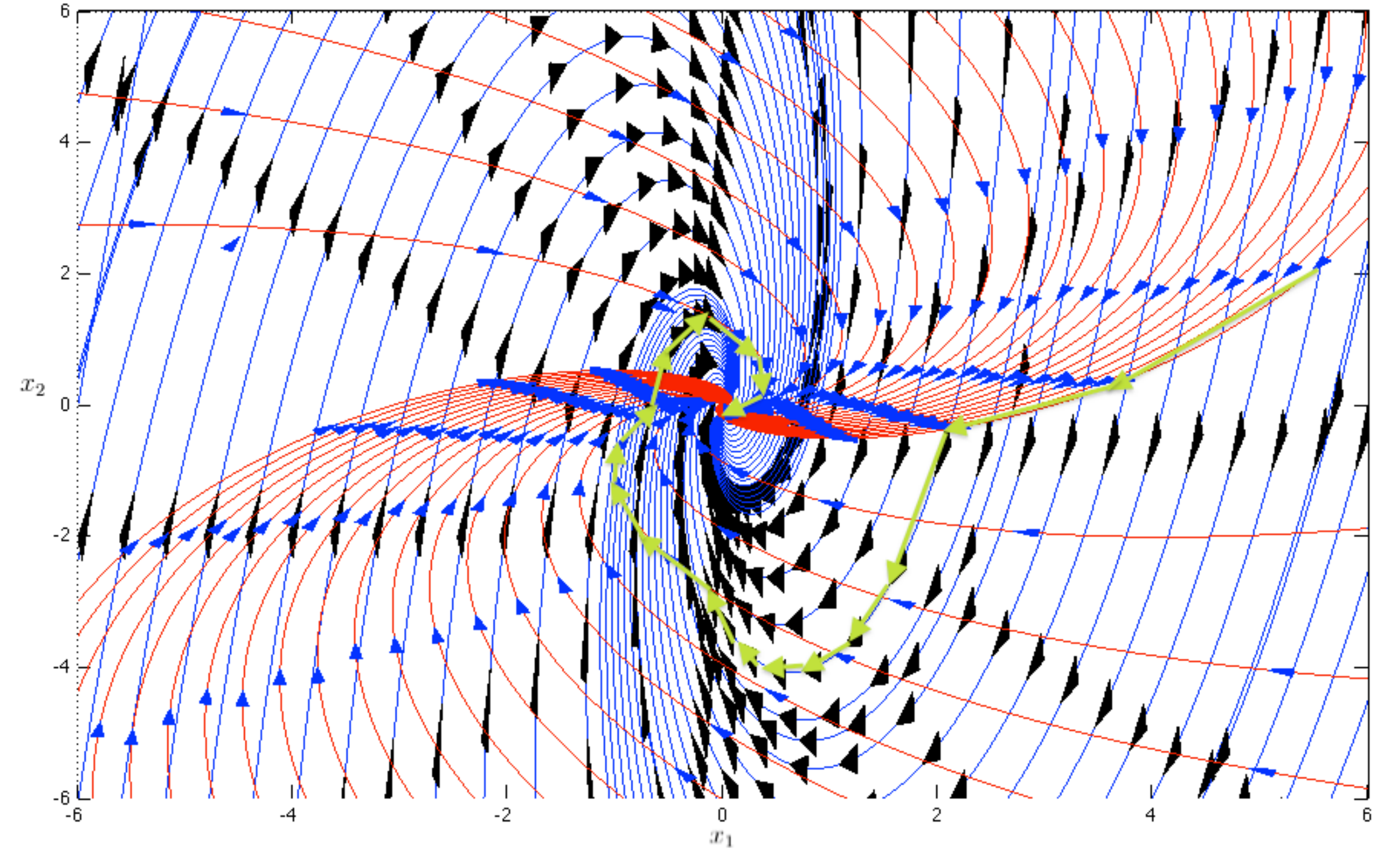} \\
($b$)\\
   \includegraphics[width=0.4\textwidth] {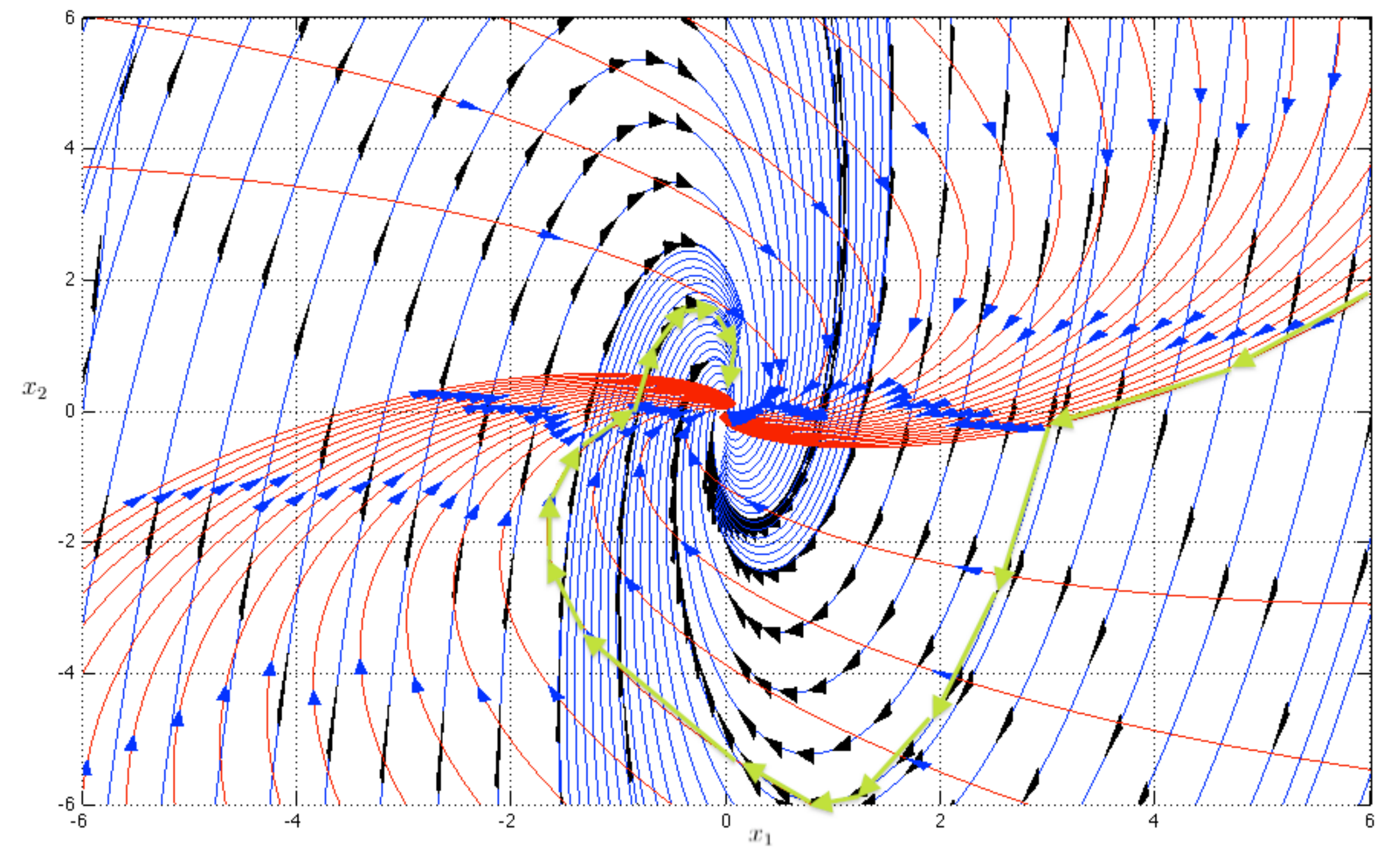}\\
   ($c$)\\
     \includegraphics[width=0.4\textwidth] {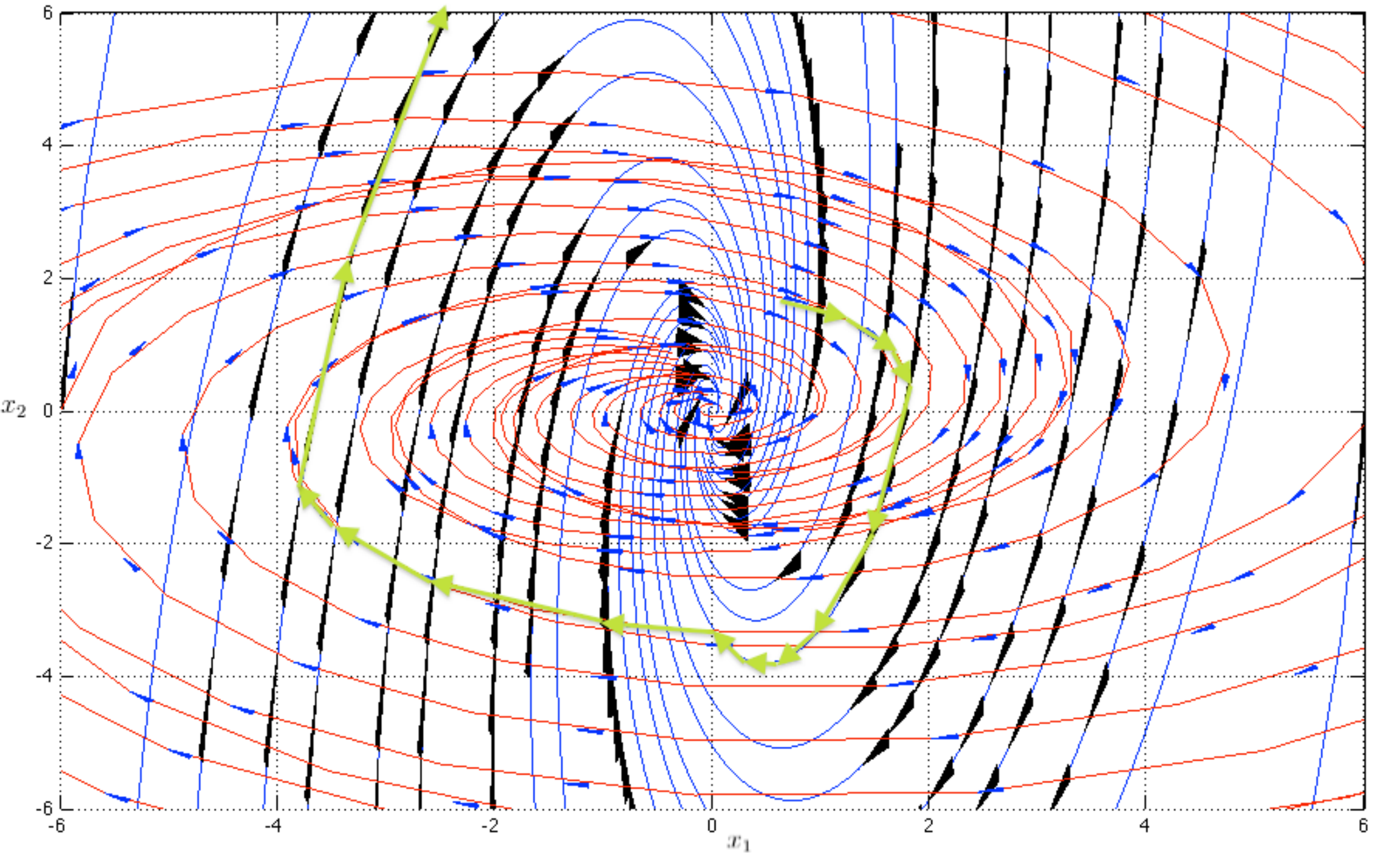}\\ 
\end{tabular}
\caption{Phase portrait of system in Example~\ref{expg} when: (a) $\alpha=0.6$ (b) $\alpha=0.8$ (c) $\alpha=0.9$. The blue trajectories refer to subsystem $1$, whereas the red ones correspond to subsystem $2$}
\label{PP0}
\end{figure}

\begin{example}
\label{expg}
Consider the switching system (\ref{FSWHM}) with $L=2$ with the following parameters:
$A_{1}=%
\begin{bmatrix}
 -0.1  &  0.1\\
   -2.0  &  -0.1
\end{bmatrix}
$, $A_{2}=%
\begin{bmatrix}
 -0.01  &  2.0\\
   -0.1  &  -0.01
\end{bmatrix}$ and order $\alpha$, $0<\alpha \leq 1$. 
\end{example}
Applying Theorem~\ref{Freq_stab_frac0}, the phase difference condition should be satisfied for all $\alpha$, $0<\alpha \leq 1$, to guarantee the stability --this condition is depicted in Fig.~\ref{Expgsw} for $0<\alpha \leq 1$ with increments of $0.1$. As can be seen, the fractional order system is stable for $\alpha\in(0,0.6]$. The phase differences when $\alpha\in\lbrack0.7,1]$ are bigger than ${\pi}/{2}$ which indicates unknown stability status, i.e., the system may be stable or unstable. For better understanding of this initial notice on the system stability, its phase portrait is shown in Fig.~\ref{PP0} for three values of $\alpha$ ($\alpha=0.6$, $\alpha=0.8$ and $\alpha=0.9$). The green trajectory is an example to show the stability or instability of the switching system. The following conclusions can be stated from these results:

\begin{itemize}
\item {When $\alpha=0.6$, it can be observed that the system is stable for arbitrary switching. This can be also confirmed by the fact that a matrix $P\begin{bmatrix}
1  &  0.2\\
   0.2  &  1
\end{bmatrix},$
satisfies the stability conditions as follows:
\small
\begin{eqnarray*}
\left(  -\left(  -A_{1}\right)  ^{\frac{1}{1.4}}\right)  ^{T}P+P\left(-\left( -A_{1}\right)  ^{\frac{1}{1.4}}\right)  = \begin{bmatrix}
-1.47 &  -0.65 \\
   -0.89 &  -0.54
\end{bmatrix} <0, \\
\left(  -\left(  -A_{2}\right)  ^{\frac{1}{1.4}}\right)  ^{T}P+P\left(-\left( -A_{2}\right)  ^{\frac{1}{1.4}}\right)  = \begin{bmatrix}
-1.45    & 0.59 \\
    0.59  &  -0.47
\end{bmatrix} <0.
\end{eqnarray*}
\normalsize
The switching region is shown in Fig.~\ref{SZ}, in which $C_1$ refers to the zone which only subsystem 1 is active, whereas $C_2$ is the zone which corresponds to subsystem 2. $D$ is a common region with a random layer where both system can be active. The red lines indicate the switching from subsystem 1 to subsystem 2, whereas the blue lines show the switching in contrary.}

\item {When $\alpha=0.8$, its phase portrait shows almost the same behaviour as with order $\alpha=0.6$. Although one cannot find a trajectory which leads to unstable switching system, the stability of the system under arbitrary switching is on doubt.}

\item {Finally, in the case of $\alpha=0.9$, the system will be unstable if it switches like the green trajectory shown in Fig.~\ref{PP0}(c).}
\end{itemize}

\begin{figure}[ptbh]
\begin{center}
\includegraphics[width=0.5\textwidth]{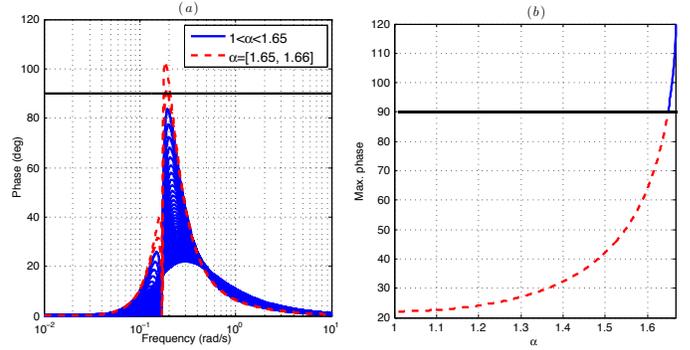}
\end{center}
\caption{Stability of the system in Example \ref{ex2} for different values of its order $\alpha$,  $1<\alpha \leq 2$: (a) Phase difference of condition (\ref{exp3eq}) (b) Maximum value of (\ref{exp3eq}) versus $\alpha$}
\label{figb1_max}%
\end{figure}

\begin{example}
\label{ex2} 
Now, consider the switching system given by (\ref{FSWHM}) with $L=2$ with the following parameters:
$A_{1}=%
\begin{bmatrix}
 -0.2 &  -1.0\\
    0.01 &   -0.1
\end{bmatrix}
$, $A_{2}=%
\begin{bmatrix}
-0.3 & 0.01 \\
   -1.0  &   -0.1
\end{bmatrix}
$ and order $\alpha$, $1<\alpha<2$.
\end{example}
It is easy to find that the subsystem $1$ is stable for when $\alpha\in(1,1.67)$, whereas the subsystem $2$ is stable for all values of $\alpha\in(1,2)$. Therefore, applying Theorem \ref{Freq_stab_frac} when $\alpha\in(1,1.67)$, the following condition
\small
\begin{align}
\left|  \arg\left( \det\left(
\begin{bmatrix}
0.03-\omega^{2}+j0.4\omega\sin \phi & 0.3+j2\omega\sin \phi\\
-0.003-j0.02\omega\sin \phi & -\omega^{2}+j0.2\omega\sin \phi)%
\end{bmatrix}
\right)  \right) - \right. \nonumber\\
\left.  \arg\left( \det\left(
\begin{bmatrix}
0.08-\omega^{2}+j0.6\omega\sin \phi & -0.004-j0.02\omega\sin \phi\\
0.4+j2\omega\sin \phi & -\omega^{2}+j0.2\omega\sin \phi)%
\end{bmatrix}
\right)  \right) \right| \nonumber\\ < \frac{\pi}{2}, \forall\omega
\label{exp3eq}%
\end{align}
\normalsize
should be satisfied, $\forall$ $\alpha$, $1<\alpha<1.67$. The phase difference (\ref{exp3eq}) is depicted in Fig.~\ref{figb1_max}(a). In order to make the results clearer, the maximum values of (\ref{exp3eq}) are also plotted in  Fig.~\ref{figb1_max}(b) versus the order of the system. It can be seen that the system is stable if $\alpha\in(1,1.65)$. The stability of the system when $\alpha\in[1.65,1.67)$ is unknown. 

\begin{figure}[ptbh]
\begin{center}
\includegraphics[width=0.4\textwidth]{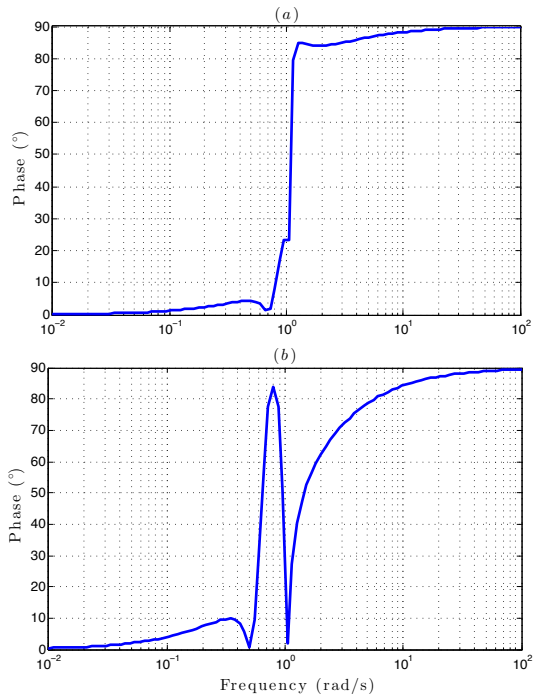}
\end{center}
\caption{Phase equivalence of H$_\beta$ in Example \ref{Rest_exp_stab}: ($a$) Applying FCI ($b$) Applying FORE}
\label{HB_FCIexp2}%
\end{figure}

\begin{example}
\label{Rest_exp_stab}
Let us consider the same feedback system as in \cite{hollot2001} with the system, the base controller and reset controller transfer functions
$ P(s)=\frac{1}{s^2+0.2s}$, $C(s)=s+1$ and, $R(s)=\frac{1}{s^\alpha+b}$,
respectively. Stability analysis of CI ($b=0, \alpha=1$), FORE ($b\neq0, \alpha=1$) and FCI ($b=0, \alpha=0.5$) will analyzed in this example.
\end{example}
The system stability will be analyzed for the first order reset element (FORE) controller, the Clegg integrator (CI) and the fractional Clegg integrator (FCI). For FORE controller, the integer-order closed-loop system can be given by:
\small
\begin{eqnarray*}
\left\{
\begin{matrix}
\dot{x}(t)=A_{cl}x=\begin{bmatrix}
0 & 1 & 0\\
0 & -0.2 & 1\\
 -1 & -1 & -b 
\end{bmatrix}x(t)\\ 
x(t^+)=A_Rx=\begin{bmatrix}
1 & 0 & 0\\
0 & 1 & 0\\
 0 & 0 & 0 
\end{bmatrix}x(t)\\ 
y=C_{cl}x=\begin{bmatrix}
1 & 1 & 0
\end{bmatrix}x(t)
\end{matrix}
\right.
\end{eqnarray*}
\normalsize
where $x(t)=\left[x_{p_1}(t),x_{p_2}(t),x_r(t)\right]^T$. And, the closed-loop system using FCI can be stated as
\small
\begin{eqnarray*}
\left\{
\begin{matrix}
D^{0.5}\mathcal{X}(t)=\mathbf{A} _{cl}\mathcal{X}(t)=\begin{bmatrix}
0 & 1 & 0 & 0 & 0\\
0 & 0 & 1 & 0 & 0\\
0 & 0 & 0 & 1 & 0\\
0 & 0 & -0.2 & 0 & 1\\
 -1 & 0 & -1 & 0 & 0
\end{bmatrix}\mathcal{X}(t)\\ 
\mathcal{X}(t^+)=\mathbf{A}_R\mathcal{X}(t)=\begin{bmatrix}
I_4 & 0_{4,1} \\
 0_{1,4} & 0 
\end{bmatrix}\mathcal{X}(t)\\ 
y=\mathbf{C}_{cl}\mathcal{X}(t)=\begin{bmatrix}
 1 & 0 & 1 & 0 & 0
\end{bmatrix}\mathcal{X}(t)
\end{matrix}
\right.
\end{eqnarray*}
\normalsize
where $\mathcal{X}(t)=\left[\mathcal{X}_{p_1}(t),\cdots,\mathcal{X}_{p_4}(t),x_r(t)\right]^T$, $\mathcal{X}_{p_1}(t)=x_{p_1}(t)$, $\mathcal{X}_{p_3}(t)=x_{p_2}(t)$. According to condition (\ref{Hb}), H$_\beta$ corresponding to FORE and FCI are  simply given by (for both case FORE and FCI $n_{\mathcal{R}}=1$ and then $P_{\mathcal{R}}=1$):
\small
\begin{equation*}
\nonumber
H^{FORE}_\beta(s)= \begin{bmatrix}
\beta & 0 & 1
\end{bmatrix}\left ( sI-A_{cl} \right )^{-1}\begin{bmatrix}
0\\ 
0\\ 
1
\end{bmatrix}=
\end{equation*}
\begin{equation}
\frac{s^2+0.2s+0.8\beta}{s^3+(b+0.2)s^2+(1+0.2b)s+1},
\label{HB_FORE_CI}
\end{equation}
\normalsize
and
\small
\begin{equation}
H^{FCI}_\beta(s)=
\begin{bmatrix}
\beta & 0 & \beta & 0 & 1
\end{bmatrix}\left ( sI-\left( -\left(-\mathbf{A}_{cl}\right)  ^{\frac{2}{3}}\right) \right )^{-1}\begin{bmatrix}
0\\ 
0\\ 
0\\ 
0\\ 
1
\end{bmatrix}.
\label{HB_FCI}
\end{equation}
\normalsize
Therefore, using Theorem \ref{FRQS}, the closed-loop systems controlled by FORE and FCI are asymptotically stable if H$^{FORE}_\beta(s)$ and H$^{FCI}_\beta(s)$ are SPR. Substituting $b=1$ in (\ref{HB_FORE_CI}), the FORE reset system is asymptotically stable for all $0.42<\beta \leq1.46$. With respect to CI (similarly to FORE with $b=0$), stability cannot be guaranteed with this theorem. And applying FCI, it can be easily stated that the system is asymptotically stable for $\beta \leq 0.62$. In addition, the phase equivalences corresponding to (\ref{HB_FORE_CI}) and (\ref{HB_FCI}) are shown in Fig. \ref{HB_FCIexp2} for $\beta=0.5$ and $b=1$. It can be seen that both phases verifies condition (\ref{FrSPR}), which has concordance with the theoretical results.

\section{Conclusions}
\label{sec_conclu}

This paper has addressed stability for two classes of hybrid systems (HS), i.e., switching and reset control systems, with fractional order dynamics by means of a Lyapunov like method. More precisely, common Lyapunov method as well as its frequency domain equivalence were generalized for the fractional order switching systems. Likewise, H$_{\beta}$-condition was used --frequency domain equivalence of Lyapunov like method for reset control systems-- to prove stability for fractional order reset control systems. The applicability and efficiency of the proposed methods were shown by some illustrative examples.

\section*{Acknowledgement}
This work has been supported by the Spanish Ministry of Economy and Competitiveness under the project DPI2012-37062-C02-02.

\bibliographystyle{IEEEtran}
\bibliography{Bibliography}

\end{document}